\documentclass[useAMS,usenatbib]{mn2e}
\usepackage{graphicx}

\newcommand{\ms}{$\,$M$_\mathrm{\odot}$}

\newcommand{\be}{\begin{equation}}
\newcommand{\ee}{\end{equation}}
\newcommand{\stars}{{\sc stars}}
\newcommand{\el}[2]{\ensuremath{^{#1}\mathrm{#2}}}
\newcommand{\ap}{\ensuremath{\mathrm{(\alpha,p)}}}
\newcommand{\pg}{\ensuremath{\mathrm{(p,\gamma)}}}
\newcommand{\pa}{\ensuremath{\mathrm{(p,\alpha)}}}
\newcommand{\ag}{\ensuremath{\mathrm{(\alpha,\gamma)}}}
\newcommand{\an}{\ensuremath{\mathrm{(\alpha,n)}}}
\newcommand{\np}{\ensuremath{\mathrm{(n,p)}}}
\newcommand{\ngam}{\ensuremath{\mathrm{(n,\gamma)}}}
\newcommand{\na}{\ensuremath{\mathrm{(n,\alpha)}}}
\newcommand{\neon}{\ensuremath{^{22}\mathrm{Ne}}}

\title[The effect of the \el{19}{F}\ap\el{22}{Ne} reaction rate 
uncertainty on the yield of fluorine from Wolf-Rayet Stars]{The effect of the \el{19}{F}\ap\el{22}{Ne} reaction rate uncertainty on the yield of fluorine from Wolf-Rayet Stars}
\author[R.J. Stancliffe, M. Lugaro, C. Ugalde et al.]{Richard J. Stancliffe\thanks{E-mail:
rs@ast.cam.ac.uk}$^1$, Maria Lugaro$^1$, Claudio Ugalde$^2$, \newauthor Christopher A. Tout$^1$, Joachim G\"orres$^2$ and Michael Wiescher$^2$\\
$^1$Institute of Astronomy, The Observatories, Madingley Road, Cambridge CB3 0HA, U.K.\\
$^2$Department of Physics and The Joint Institute for Nuclear Astrophysics,
University of Notre Dame, Notre Dame,\\ Indiana 46556, U.S.A.}
\begin{document}
\bibliographystyle{mn2e}

\date{Accepted 0000 December 00. Received 0000 December 00; in original form 0000 October 00}

\pagerange{\pageref{firstpage}--\pageref{lastpage}} \pubyear{0000}

\maketitle

\label{firstpage}

\begin{abstract}

In the light of recent recalculations of the \el{19}{F}\ap\el{22}{Ne} reaction rate we present results of the
expected yield of \el{19}{F} from Wolf-Rayet (WR) stars. We have computed models using the upper and
lower limits for the rate in addition to the recommended rate and hence we constrain the uncertainty in
the yield with respect to this reaction. We find a yield of $3.1 \times10^{-4}$\ms\ of \el{19}{F} with
our recommended rate and a difference of a factor of two between the yields computed with the upper and
lower limits. 
In comparison with previous work we find a difference in the yield of
approximately a factor of 4, connected to a different choice of mass loss. Model uncertainties must be carefully evaluated in order to obtain a reliable estimate of the yield of fluorine from WR stars together with its uncertainties.
\end{abstract}


\section{Introduction}

The production site of \el{19}{F} has been a major
puzzle for nucleosynthesis for a long time. It was predicted by 
\citet{Goriely89} that \el{19}{F} should be manufactured in asymptotic giant branch stars, and these are 
currently the only observationally confirmed site for 
fluorine production \citep{1992A&A...261..164J}. 
Other sites and mechanisms for the galactic production of fluorine have been proposed. The 
neutrino process operating during type-II supernova explosions can
produce fluorine \citep{1995ApJS..101..181W}. Moreover, fluorine can be synthesised
during core He-burning and ejected via the strong winds of Wolf-Rayet (WR)
stars \citep{1993nuco.conf..487M,2000A&A...355..176M}. It appears that the 
contributions of asymptotic giant branch and WR stars have to be included 
in the computation of the chemical evolution of the Galaxy to account for 
the observations of fluorine in the Milky Way \citep{2004MNRAS.354..575R}. 

The uncertainties of the nuclear reaction rates involved in the production and destruction of \el{19}{F} in the case of asymptotic 
giant branch stars have been studied in detail by \citet{2004ApJ...615..934L}. 
Here we examine fluorine production in WR stars with regard to the uncertainties of the
\el{19}{F}\ap\el{22}{Ne} rate.

Fluorine production in WR stars is of secondary nature because it relies on the presence of \el{13}{C} and
\el{14}{N} produced during H burning. During core helium burning the production mechanism for \el{19}{F}
is as follows. The seed \el{14}{N} undergoes $\alpha$ capture to produce the unstable nucleus 
\el{18}{F}
which $\beta$-decays to \el{18}{O}. In the presence of protons the reaction \el{18}{O}\pa\el{15}{N} can
occur which then leads to the formation of \el{19}{F} via the reaction \el{15}{N}\ag\el{19}{F}.
Protons may be present in a helium burning core from the reaction \el{14}{N}\np\el{14}{C}, with the
neutrons coming from the reaction \el{13}{C}\an\el{16}{O}. The important destruction mechanism
associated with the core helium burning phase is the reaction \el{19}{F}\ap\el{22}{Ne}. New
estimates for the rate of this reaction have been presented by \citet{2004ApJ...615..934L}. It is very
difficult to gather experimental data at low energies so the estimated uncertainties are
of fourteen orders of magnitudes at a temperature of about $2 \times 10^8$ K, typical of core helium
burning. Thus it is necessary to investigate the impact of such a large uncertainty 
on the yield of fluorine from WR stars.

In section~\ref{sec:code} we describe the details of the evolution code used and its nucleosynthesis
routine. We briefly review the details of the \el{19}{F}\ap\el{22}{Ne} reaction in section~\ref{sec:f19ap}. In section~\ref{sec:res} we present the results of the simulations. Conclusions and 
directions for further work are outlined in section~\ref{sec:conclusions}.

\section{The stellar evolution code}
\label{sec:code}

Calculations were made using the evolution code \stars\ originally developed by
\citet{1971MNRAS.151..351E} and updated by many contributors
\citep[e.g.][]{1995MNRAS.274..964P}. The current version of the code 
employs a fully simultaneous solution of the equations of stellar structure, 
nuclear burning and mixing \citep*{2004MNRAS.352..984S}. Opacities are taken from \citet{1992ApJS...79..507R} and \citet{1994ApJ...437..879A}. The evolution code uses an approximate reaction network comprising only those elements -- namely \el{1}{H}, \el{4}{He}, \el{12}{C}, \el{14}{N}, \el{16}{O} and \el{20}{Ne} -- whose reactions are energetic enough to affect the stellar structure.

In order to fully investigate the nucleosynthesis that takes place in stars we
require an additional routine that comprises a full network for the elements that
we are interested in. Within the framework of the \stars\ code, this takes the
form of a subroutine called after the convergence of a model with the main code. 
The nucleosynthesis subroutine takes the structural details from this converged model and uses them to 
compute the nucleosynthesis of all the elements we are interested in. 

\subsection{The nucleosynthesis routine}

The nucleosynthesis routine deals with 40 isotopes. These
include all stable particles, and a few unstable ones, from neutrons and
deuterium up to \el{34}{S} and some isotopes from the iron group. It also
computes its own values for the isotopes included in the 
evolutionary code as the compositions from the evolutionary part of the 
code will deviate slightly from those of the nucleosynthesis part. 
A list of all the isotopes included in the nucleosynthesis code is given 
in Table~\ref{tab:elements}.

\begin{table}
\begin{center}
\begin{tabular}{c}
Light isotopes \\ \hline
{\bf $^\mathbf{1}$H},$n$,\el{2}{H},\el{3}{He},{\bf $^\mathbf{4}$He},\el{7}{Li},\el{7}{Be},\el{11}{B},{\bf $^\mathbf{12}$C},\el{13}{C}, \\
$\mathit{^{14}C}$,{\bf $^\mathbf{14}$N},\el{15}{N},{\bf $^\mathbf{16}$O},\el{17}{O},\el{18}{O},\el{19}{F},{\bf $^\mathbf{20}$Ne},\el{21}{Ne},\el{22}{Ne}, \\
$\mathit{^{22}Na}$,\el{23}{Na},\el{24}{Mg},\el{25}{Mg},\el{26}{Mg},$\mathit{^{26}Al^m}$,$\mathit{^{26}Al^g}$,\el{27}{Al},\el{28}{Si},\el{29}{Si}, \\
\el{30}{Si},\el{31}{P},\el{32}{S},\el{33}{S},\el{34}{S}\\
\hline
Iron group isotopes \\ \hline
\el{56}{Fe},\el{57}{Fe},\el{58}{Fe},$\mathit{^{59}Fe}$,$\mathit{^{60}Fe}$ \\
\el{59}{Co},\el{58}{Ni},$\mathit{^{59}Ni}$,\el{60}{Ni},\el{61}{Ni} \\ \hline
\end{tabular}
\end{center}
\caption{Isotopes included in the nucleosynthesis code. Isotopes also included in the structural part of the code are highlighted in bold. Unstable isotopes are in italics.}
\label{tab:elements}
\end{table}

Unstable nuclei that are not included in the network 
are treated as if their decay were instantaneous. This approximation is fair for 
all light isotopes with half-lives from seconds to hours in conditions of 
core helium burning. 
For the unstable isotopes considered in the network, the decay lifetimes are the terrestrial values
given by \citet{krane}.

\subsubsection{Charged particle reaction rates}

In order to couple the nucleosynthesis network 63 charged particle
reactions are required. The rates are taken from a variety of sources and
are listed in Table~\ref{tab:p} (proton captures)
and Table~\ref{tab:alpha} ($\alpha$ captures). The rate of the reaction
\el{3}{He}(\el{3}{He},2p)\el{4}{He} is that given by 
\citet{1988ADNDT..40..283C}, as are the rates for carbon and oxygen burning
reactions.

The nucleosynthesis routines were designed to employ the ready-to-use fits to the reaction rates from the REACLIB library \citep[1991 updated version of][]{1986ana..work..525T}, updated where possible to include the latest experimental results \citep[see][~for full details]{2004ApJ...615..934L}. For some of the rates involved in the production of \el{19}{F}, such as \el{15}{N}\ag\el{19}{F}, the rates are virtually the same as those presented in the NACRE compilation \citep{1999NuPhA.656....3A}. In other cases, such as the rates \el{14}{N}\ag\el{18}{F} and \el{18}{O}\ag\el{22}{Ne}, the rates used are updates with respect to the NACRE rates.

\begin{table}
\begin{center}
\begin{tabular}{cc}
Reaction & Source \\ \hline
\el{1}{H}(p,$\beta^+\nu$)\el{2}{H} & REACLIB \\
\el{2}{H}\pg\el{3}{He} & CF88 \\
\el{7}{Li}\pa\el{4}{He} & CF88 \\
\el{7}{Be}\pg2\el{4}{He} & CF88 \\
\el{12}{C}\pg\el{13}{N} & CF88 \\
\el{13}{C}\pg\el{14}{N} & NACRE \\
\el{14}{C}\pg\el{15}{N} & W90 \\
\el{14}{N}\pg\el{15}{O} & CF88 \\
\el{15}{N}\pg\el{16}{O} & CF88 \\
\el{15}{N}\pa\el{12}{C} & CF88 \\
\el{16}{O}\pg\el{17}{F} & CF88 \\
\el{17}{O}\pg\el{18}{F} & L90,B95 \\
\el{18}{O}\pg\el{19}{F} & CF88 \\
\el{18}{O}\pa\el{15}{N} & CF88 \\
\el{19}{F}\pg\el{20}{Ne} & CF88 \\
\el{19}{F}\pa\el{16}{O} & CF88 \\
\el{21}{Ne}\pg\el{22}{Na} & EL95 \\
\el{22}{Ne}\pg\el{23}{Na} & EL95 \\
\el{22}{Na}\pg\el{23}{Mg} & SC95,ST96 \\
\el{23}{Na}\pg\el{24}{Mg} & EL95 \\
\el{23}{Na}\pa\el{20}{Ne} & EL95 \\
\el{24}{Mg}\pg\el{25}{Al} & 99tDC \\
\el{25}{Mg}\pg\el{26}{Al}$^g$ & I96 \\
\el{25}{Mg}\pg\el{26}{Al}$^m$ & I96 \\
\el{26}{Mg}\pg\el{27}{Al} & I90 \\
\el{27}{Al}\pg\el{28}{Si} & CF88 \\
\el{27}{Al}\pa\el{24}{Mg} & T88,C88 \\
\el{28}{Si}\pg\el{29}{P} & G90 \\
\el{29}{Si}\pg\el{30}{P} & CF88 \\
\el{30}{Si}\pg\el{31}{P} & CF88 \\
\hline
\end{tabular}
\end{center}

\caption{Proton capture reactions and the sources from which their rates were
taken. Key: C88 \citep{1988NuPhA.487..433C}, CF88 \citep{1988ADNDT..40..283C},
B95 \citep{1995PhRvL..74.2642B}, EL95 \citep{1995ApJ...451..298E}, G90
\citep{1990NuPhA.517..329G}, I90 \citep{1990NuPhA.512..509I}, I96
\citep{1996PhRvC..53..475I}, L90 \citep{1990A&A...240...85L}, NACRE
\citep{1999NuPhA.656....3A}, SC95 \citep{1995NuPhA.591..227S}, ST96
\citep{1996NuPhA.601..168S}, REACLIB (1991 updated version of \citealt{1986ana..work..525T}), T88 \citep{1988NuPhA.477..105T}, W90 \citep{1990ApJ...363..340W}.}

\label{tab:p}
\end{table}

\begin{table}
\begin{center}
\begin{tabular}{cc}
Reaction & Source \\ \hline
\el{4}{He}($\alpha\alpha,\gamma$)\el{12}{C} & CF88 \\
\el{7}{Li}\ag{11}{B} & CF88 \\
\el{12}{C}\ag\el{16}{O} & CF88 \\
\el{13}{C}\an\el{16}{O} & D95 \\
\el{14}{C}\ag\el{18}{O} & JG01 \\
\el{14}{N}\ag\el{18}{F} & G00 \\
\el{15}{N}\ag\el{19}{F} & deO96 \\
\el{16}{O}\ag\el{20}{Ne} & CF88 \\
\el{17}{O}\an\el{20}{Ne} & D95 \\
\el{18}{O}\ag\el{22}{Ne} & D03 \\
\el{18}{O}\an\el{21}{Ne} & D95 \\
\el{19}{F}\ap\el{22}{Ne} & U04 \\
\el{20}{Ne}\ag\el{24}{Mg} & CF88 \\
\el{21}{Ne}\ag\el{25}{Mg} & CF88 \\
\el{21}{Ne}\an\el{24}{Mg} & D95 \\
\el{22}{Ne}\ag\el{26}{Mg} & K94 \\
\el{22}{Ne}\an\el{25}{Mg} & K94 \\
\el{23}{Na}\an\el{26}{Al}$^g$ & CF88 \\
\el{23}{Na}\an\el{26}{Al}$^m$ & CF88 \\
\el{24}{Mg}\ag\el{28}{Si} & CF88 \\
\el{25}{Mg}\ag\el{29}{Si} & CF88 \\
\el{25}{Mg}\an\el{28}{Si} & CF88 \\
\el{25}{Mg}\ap\el{28}{Al} & CF88 \\
\el{26}{Mg}\ag\el{30}{Si} & CF88 \\
\el{26}{Mg}\an\el{29}{Si} & CF88 \\
\el{26}{Mg}\ap\el{29}{Al} & CF88 \\
\el{27}{Al}\ag\el{31}{Si} & REACLIB \\ \hline
\end{tabular}
\end{center}

\caption{Reactions involving $\alpha$ capture and the sources from which their rates were taken. Key:
CF88 \citep{1988ADNDT..40..283C}, D95 \citep{1995AIPC..327..255D}, deO96 \citep{1996NuPhA.597..231D},
G00 \citep{2000PhRvC..62e5801G}, JG01 \citep{2001NuPhA.688..508J}, D03 \citep{2003PhRvC..68b5801D}, K94
\citep{1994ApJ...437..396K}, REACLIB, 1991 updated version of \citet{1986ana..work..525T}, U04
\citep{Ugalde}.}
\label{tab:alpha}

\end{table}

\subsubsection{Neutron capture rates} 
\label{sec:neutron}

A total of 45 neutron capture reactions are required for the network. The work of
\citet{2000ADNDT..76...70B} was used as the main source. 
Supplementary \ngam\ data are taken from \citet{2000ADNDT..75....1R}
for captures by \el{59,60}{Fe}. Rates for the reaction \el{33}{S}\na\el{30}{Si}
were taken from \citet{1995PhRvC..51..379S}. For the reactions
\el{26}{Al}$^g$\np\el{26}{Mg} and \el{26}{Al}$^g$\na\el{23}{Na} rates are
from \citet{1997PhRvC..56.1138K}. The important reaction rate for 
\el{14}{N}\np\el{14}{C} is from \citet{1995AIPC..327..173G}, which is 
in agreement with previous experimental \citep{1989PhRvC..39.1655K} and theoretical \citep{1969ApJ...157..659B} estimates. This rate is approximately a factor of two higher than the rate proposed by \citet{brehm:88} and used by \citet{1993nuco.conf..487M,2000A&A...355..176M}.

For neutron captures by \el{59}{Ni} we take reaction rates from \citet{1976ADNDT..18..305H} and this is also the source of the rate of the reaction \el{17}{O}\na\el{14}{N}.

In addition to this, two neutron sinks are included to account for neutron
captures by those elements not included in the network. The first 
sink is emulated by the reaction \el{34}{S}\ngam\el{35}{S} and represents 
nuclei between \el{34}{S} and the iron group. The second sink 
is emulated by the reaction \el{61}{Ni}\ngam\el{62}{Ni} and represents
captures by all the heavy elements above \el{61}{Ni}. In any case these reactions 
are not very important in core He-burning conditions where the main 
neutron absorber is \el{14}{N}. 

\section{The \el{19}{F}\ap\neon\ reaction rate}
\label{sec:f19ap}

\begin{figure}
\includegraphics[width=8cm]{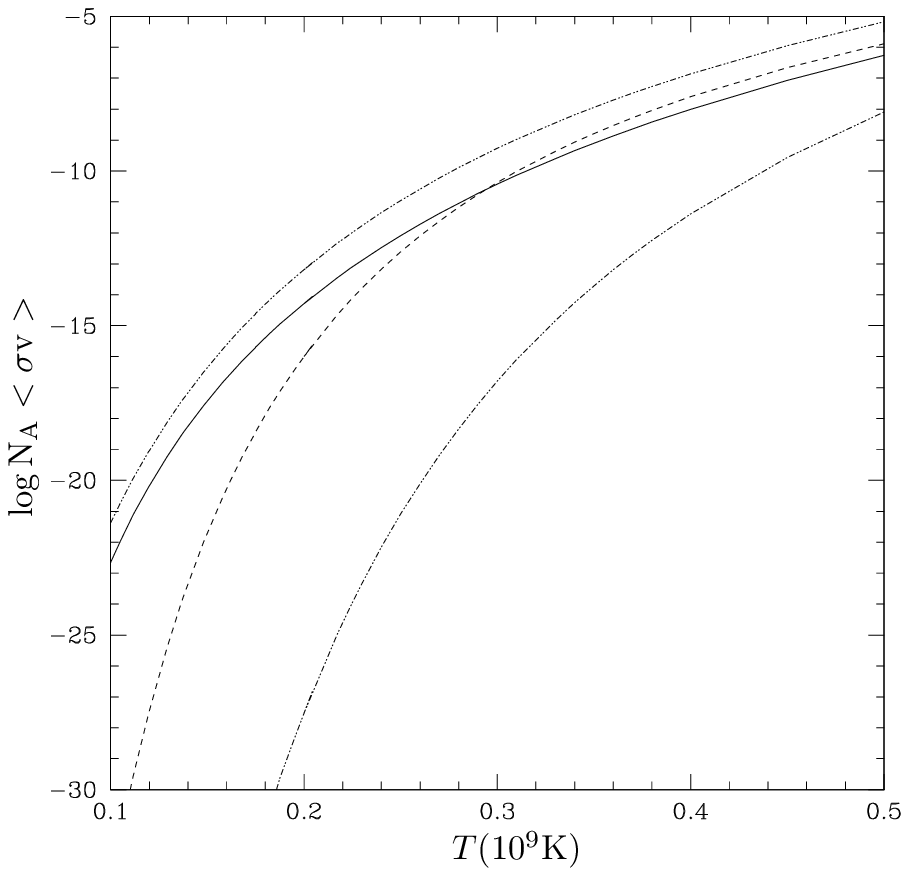}
\caption{The recommended rate of the \el{19}{F}\ap\el{22}{Ne} reaction is plotted as a solid line. 
Upper and lower limits of the rates are plotted as dotted lines, while the dashed line represents 
the \el{22}{Ne}\an\el{25}{Mg} reaction rate. 
}
\label{fig:rates}
\end{figure}

The \el{19}{F}\ap\el{22}{Ne} reaction rate is the main destruction 
channel of \el{19}{F} in WR stars. However there is very little experimental 
data available at low energies. As described in detail by 
\citet{2004ApJ...615..934L}, the level density in the
compound nucleus \el{23}{Na} has been analysed on the basis of, as yet unpublished, low-energy \el{19}{F}\ap\ resonance measurements by \citet{Ugalde}.  
We found that the level density is too low to apply the Hauser-Feshbach 
approach \citep{1997PhRvC..56.1613R}, which yields a rate in reasonable agreement 
with the estimate by \citet{1988ADNDT..40..283C}.
Thus the rate needs to be 
calculated from determination of the strengths, $\omega\gamma$, for the single
resonances. The resulting recommended rate is shown in Figure~\ref{fig:rates} 
together with the upper and lower limits.
In the temperature range of core He burning the recommended rate is more than one
order of magnitude smaller than the rate estimated by 
\citet{1988ADNDT..40..283C} and used in the previous calculations. 
The lower limit for the rate is several orders of magnitude 
smaller than the recommended rate.

\section{Results}
\label{sec:res}

To test the effects of varying the \el{19}{F}\ap\el{22}{Ne} reaction rate a
60\ms\ model of $Z=0.02$ was evolved from the pre-main sequence through to the WR
phase with the recommended reaction rate and the upper and lower limits. Initial abundances for all the elements considered are taken from \citet{1989GeCoA..53..197A}. This means we take the initial \el{19}{F} mass fraction as $4.051\times10^{-7}$. Mass loss was applied by the mass-loss rates of \citet*{1988A&AS...72..259D} 
for pre-WR evolution and the rates of \citet{1989A&A...220..135L} for the
WR phase. The switch to WR mass loss is made when the surface hydrogen abundance by mass fraction reaches 0.4 and the surface temperature exceeds $10^4$\,K. These may not be the ideal prescriptions to use \citep[see][]{2004MNRAS.353...87E} but our aim is to investigate the effect of varying the \el{19}{F}\ap\el{22}{Ne} reaction rate not the influence of mass-loss rate on WR evolution. 

\begin{figure*}
\includegraphics[width=8.5cm]{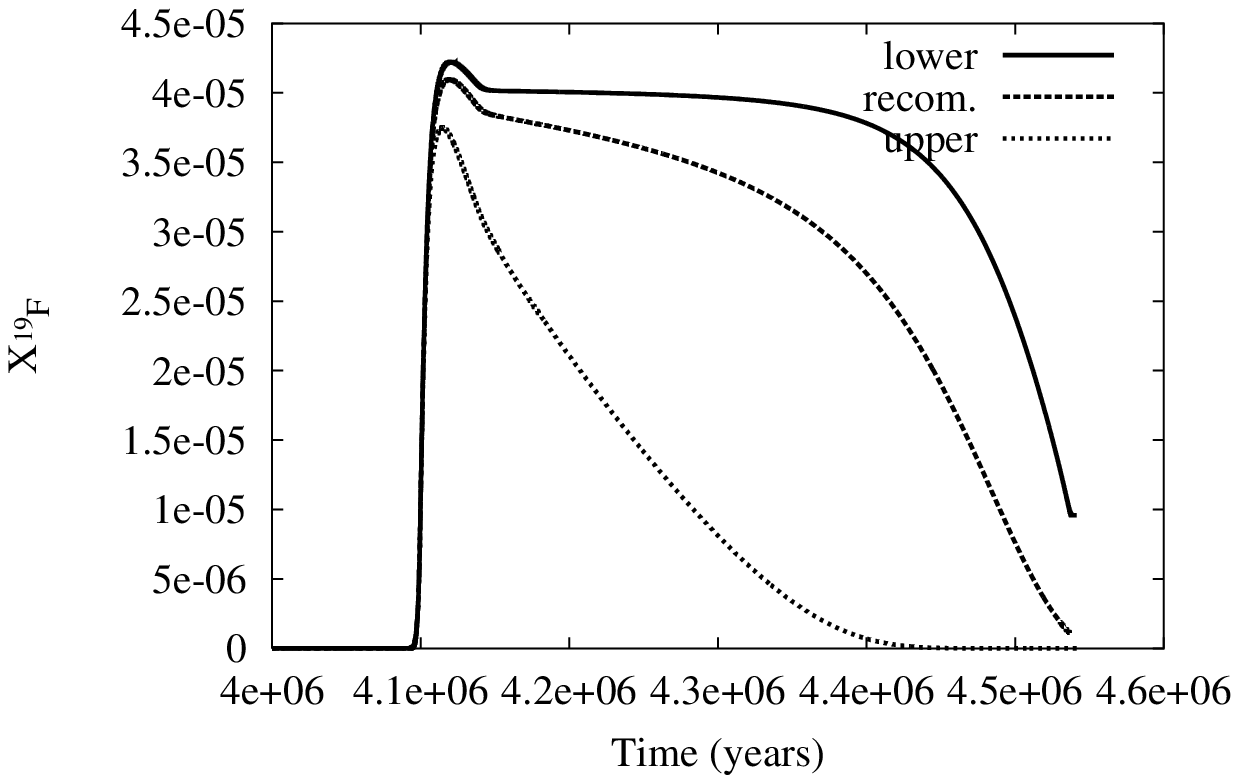}
\includegraphics[width=8.5cm]{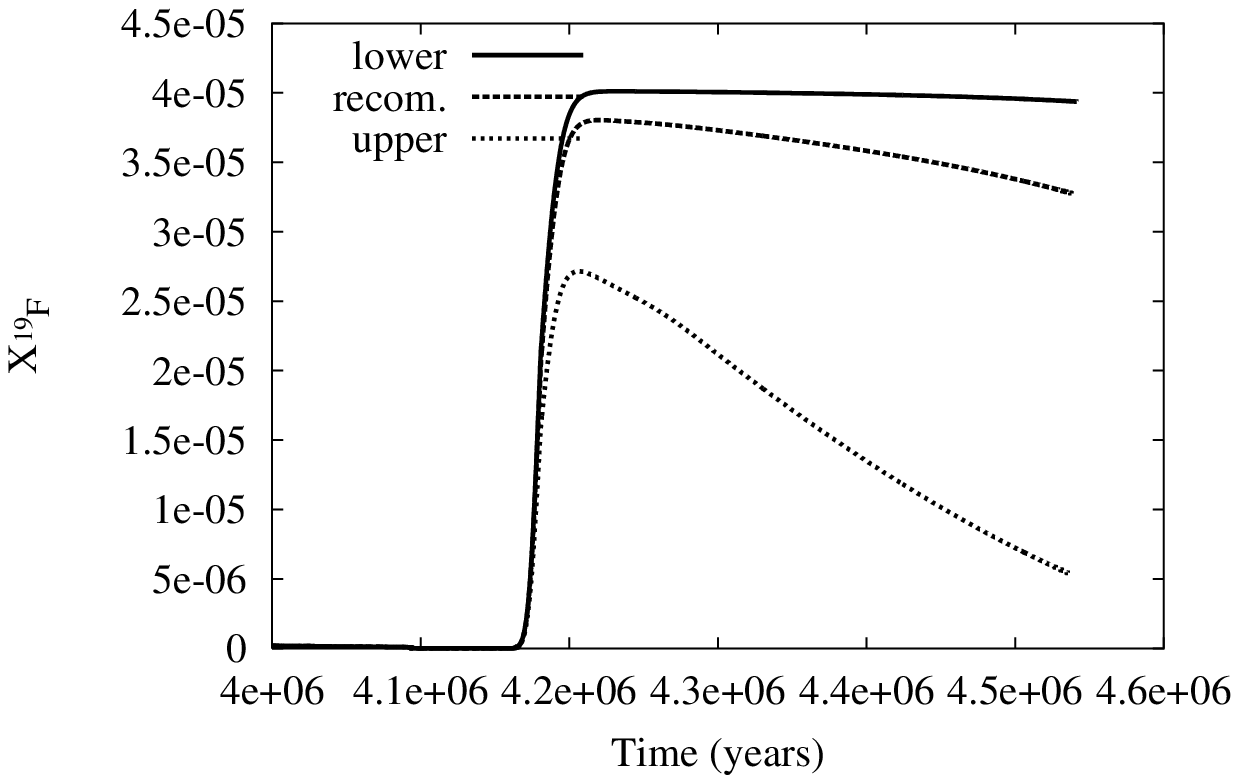}
\caption{Left panel: mass fraction of \el{19}{F} present in the core
as a function of time. The sharp increase in the mass fraction at $4.1\times10^6$ yr is the onset of core He-burning. Right panel: mass fraction of \el{19}{F}
at the stellar surface as a function of time.}
\label{fig:F19}
\end{figure*}

Figure~\ref{fig:F19} shows the mass fraction of \el{19}{F} in the core and at the surface as a function of time. Initially a rapid increase in the core fluorine
abundance occurs because all the available \el{13}{C} and \el{14}{N} undergo
$\alpha$ captures, opening the path to \el{19}{F} production. It takes about 
$8\times10^4$ yr from this time to have a notable increase in 
the abundance of \el{19}{F} at the surface when stellar winds have stripped away the stellar surface exposing the formerly convective core. Between this point and the end of the life of the star (the WC phase) approximately 9\ms\ of material is lost from the surface.

\begin{figure}
\includegraphics[width=8cm]{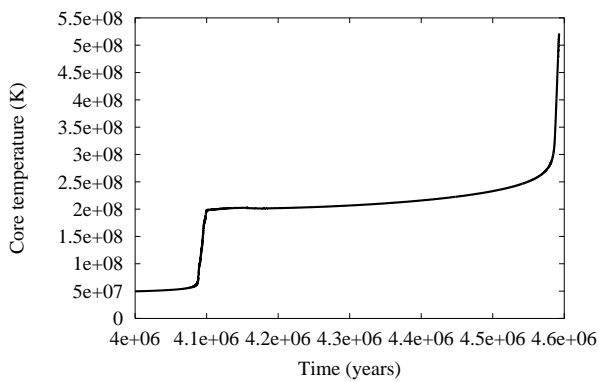}
\caption{Core temperature as a fuction of time for the He-burning phase.}
\label{fig:coreT}
\end{figure}

A plot of the core temperature as a function of time during the He-burning is shown in
Figure~\ref{fig:coreT}. In the case of the upper limit of the \el{19}{F}\ap\el{22}{Ne} reaction, we see
that temperatures are sufficiently high for the reaction to be active in the core soon after the maximum
abundance of \el{19}{F} has been reached and the fluorine abundance in the core rapidly drops from its
peak value. For the recommended rate the initial decline is much slower and becomes steeper
after $4.4\times10^6$ yr. The lower limit to the rate leads to an almost constant core abundance of
\el{19}{F} until about $4.4\times10^6$ yr later at which point fluorine is appreciably destroyed.
The destruction of fluorine in the lower-limit case, and the steeper decline in the recommended case, 
that occur at a time of about $4.4\times10^6$ yr are to be attributed to another destruction channel 
that is opening at around such a time in the centre of the star, the \el{19}{F}\ngam\ reaction where the neutrons are provided by the activation of the $^{22}$Ne($\alpha,n$)$^{25}$Mg reaction rate. The rate we use for this neutron source reaction is practically the same as that recommended by NACRE. Such a rate is comparable to the recommended limit of the \el{19}{F}\ap\el{22}{Ne} reaction and it is much higher than the lower limit for this reaction, as illustrated in Figure~\ref{fig:rates}. The estimate of the neutron capture cross section of \el{19}{F} is relatively old and needs to be revaluated \citep{2000ADNDT..76...70B} but a different rate is unlikely to have a large impact on the \el{19}{F} yield from WR stars as the surface abundance is not affected by this destruction channel, as shown by the left panel of Figure~\ref{fig:F19}.

\begin{table}
\begin{center}
\begin{tabular}{lc}
Reaction Rate & Yield ($10^{-4}$\ms) \\ \hline
Upper limit   & 1.7 \\
Recommended   & 3.1 \\
Lower limit   & 3.4 \\
\hline
\end{tabular}
\caption{The yield of \el{19}{F} obtained with the indicated 
\el{19}{F}\ap\el{22}{Ne} reaction rates.}
\label{tab:yield}
\end{center}
\end{table}

The yield of \el{19}{F} from each of the evolution runs is presented in
Table~\ref{tab:yield}. It is defined as
\be
 p = \int_0^\tau \dot{M}(t)[X_\mathrm{s}(t) - X_0]dt
\ee
where $\tau$ is the lifetime of the star, $\dot{M}(t)$ the mass-loss rate at age
$t$, $X_\mathrm{s}(t)$ the surface fluorine abundance at time $t$ and $X_0$ the initial
abundance of fluorine. The rapid destruction of fluorine that occurs with the upper limit of the \el{19}{F}\ap\el{22}{Ne} rate leads to a yield approximately a factor of two lower than the cases computed using the recommend and the lower limit for the rate. 

In order to compare our results with those presented by \citet{1993nuco.conf..487M,2000A&A...355..176M} for the equivalent
model we must consider the results of our calculation computed with the upper limit for the
rate. This is equivalent to the rate proposed by \citet{1988ADNDT..40..283C}. Our yield is very close to the value of $1.2\times10^{-4}$\ms\ calculated by \citet{1993nuco.conf..487M}, who used the standard mass-loss rate that we adopted. This yield is approximately a factor of four smaller than that calculated by \citet{2000A&A...355..176M}, who adopted an enhanced mass-loss rate. The enhanced mass loss rates used by \citet{2000A&A...355..176M} will expose
the He core sooner and so provide a higher \el{19}{F} yield. The comparison
underlines the importance of mass-loss rates on the calculations of yields
from these stars.

We also note that many reaction rates that influence the production of
fluorine have been updated in our calculations with respect to those of
\citet{1988ADNDT..40..283C} that were used by Meynet \& Arnould. Moreover, we
have not included any convective overshooting. This means that we have a
smaller convective core than Meynet \& Arnould, which affects the
luminosity of the star and thus the mass-loss rate.

\section{Conclusions}
\label{sec:conclusions}

We have calculated the yield of \el{19}{F} from a Wolf-Rayet star of 60\ms\ at
$Z=0.02$ with the upper, recommended and lower limits for the rate of the
\el{19}{F}\ap\el{22}{Ne} reaction. We find that there is a difference of 
a factor of two between the yields computed with the upper and lower 
limits for the rate. As future work the
effects of the new rates for the \el{19}{F}\ap\el{22}{Ne} reaction
need to be determined across a wide range of stellar masses and metallicities and the effect of the uncertainties associated with the other rates affecting the production of fluorine should also be analysed.
 
We also find that, for a given rate of the \el{19}{F}\ap\el{22}{Ne} reaction, 
the \el{19}{F} from Wolf-Rayet stars is reduced by a factor of about 4 with respect to previous calculations, made with an enhanced mass-loss rate \citep{2000A&A...355..176M}. These model uncertainties should be carefully evaluated.

\section{Acknowledgements}
We thank the anonymous referee for helpful remarks that have corrected and improved this paper. RJS wishes to thank PPARC for a studentship and John Eldridge for assistance with mass-loss rates. CAT thanks Churchill College for a fellowship.

\bibliography{ME1446rv}

\label{lastpage}

\end{document}